\documentclass[aps,pra,showpacs,twocolumn]{revtex4}
\usepackage{amsmath,graphicx,epsfig,amssymb,subfigure,times,dsfont}
\usepackage[usenames]{color}
\usepackage{bbold}

\newcommand {\Fig}[1] {Fig.~\ref{#1}}
\newcommand {\Eqn}[1] {Eq.~(\ref{#1})}
\newcommand {\Sec}[1] {Sec.~\ref{#1}}

\begin{document}

\title{The area law and real-space renormalization}

\author{Andrew~J.~Ferris}
\affiliation{D\'epartement de Physique, Universit\'e de Sherbrooke, Qu\'ebec, J1K 2R1, Canada}
\date{\today}

\begin{abstract}
Real-space renormalization-group techniques for quantum systems can be divided into two basic categories --- those capable of representing correlations following a simple boundary (or area) law, and those which are not. I discuss the scaling of the accuracy of gapped systems in the latter case and analyze the resultant spatial anisotropy. It is apparent that particular points in the system, that are somehow `central' in the renormalization, have local quantities that are much closer to the exact results in the thermodynamic limit than the system-wide average. Numerical results from the tree-tensor network and tensor renormalization-group approaches for the 2D transverse-field Ising model and 3D classical Ising model, respectively, clearly demonstrate this effect.
\end{abstract}


\maketitle

\section{Introduction}

Solving large, quantum mechanical systems is very challenging, primarily because the dimension of the Hilbert space describing a system with many components grows exponentially with the number of components. Direct approaches to such problems, for instance by exact diagonalization, quickly become intractable, even for relatively small 2D and 3D quantum systems. Quantum Monte Carlo (QMC) is another direct approach (exact up to statistical error), but the sign problem causes difficulties for many systems of interest --- such as fermionic, frustrated, or dynamical problems.

Thus, in order to garner meaningful information about large quantum systems, clever approximations need to be employed. In this vain, many analytic and numeric techniques have been developed over the last 80 years. In this paper, I will focus specifically on numerical real-space renormalization-group (RG) techniques, which can be applied to both quantum and classical problems.

The process of renormalization takes a divide-and-conquer approach to the problem, by tackling different parts of the system (or Hilbert space) separately or in succession, carefully simplifying or compressing the pieces at each step. In real-space RG, the renormalization procedure groups together spatial regions of the system, referred to as blocks. At each step, two neighboring blocks are combined, and then simplified (for instance, by truncating the Hilbert space). As the renormalization proceeds, the blocks include larger and larger portions of the initial system, until the entire system of interest is encapsulated. If the RG reaches a fixed-point, we can say we have reached the thermodynamic limit.

\begin{figure}
\begin{centering}
 \includegraphics[width=0.85\columnwidth]{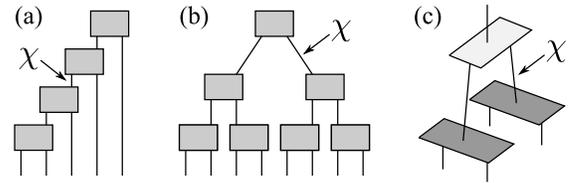}
  \caption{(a) Depiction of an MPS (in the unitary gauge) pictured in terms of renormalized Hilbert spaces. Each tensor adds a physical site (bottom) and passes the Hilbert space (truncated to $\chi$) upwards. (b) The 1D TTN for 8 sites. Each tensor combines and renormalizes two neighboring blocks. (c) A single layer of the 2D TTN, where a 2$\times$2 square is renormalized into a single site, first by combining in the $x$ then $y$ directions. \label{fig:TNs}}
\end{centering}
\end{figure}

One of the most successful real-space RG algorithms is the density-matrix renormalization group (DMRG)~\cite{White1992}, which very accurately describes (quasi-)1D systems. In this approach, a single site is added to the block at each step, and the optimal Hilbert space (limited to some dimension $\chi$) for the combined system is determined. Working in reverse, the procedure generates a variational wave-function known as a matrix-product state (MPS) (see \Fig{fig:TNs}~(a))~\cite{Oestlund1995a}. DMRG can then be thought of as an optimization algorithm that sweeps over the tensors in the MPS, targeting the state with lowest energy.

A similar approach is the tree-tensor network (TTN)~\cite{Shi2006}, in which neighboring blocks are successively combined, depicted in \Fig{fig:TNs}~(b). In this direct coarse-graining approach, the physical volume of each block doubles at each step. This ansatz is used less frequently than MPS/DMRG because the numerical cost is higher for a given amount of system entanglement or accuracy ($\mathcal{O}(\chi^4)$ vs. $\mathcal{O}(\chi^3)$). The tensor network can easily be extended to higher-dimensional systems (see \Fig{fig:TNs}~(c))~\cite{Tagliacozzo2009,Murg2010,Li2012}.

The major problem of using the TTN (or MPS/DMRG~\cite{White1998}) in two- or higher-dimensions is that they do not respect the area-law for entanglement entropy with fixed $\chi$~\cite{Tagliacozzo2009,Stoudenmire2012}. Generally speaking, gapped phases of local Hamiltonians are expected to have `local' correlations. Thus, the amount of entanglement between a large, contiguous block and the rest of the system should scale proportional to the boundary area separating the regions. On the other hand, wave-functions generated by MPS or TTN contain arbitrary large blocks with bounded entanglement (depending on $\chi$), and therefore have poor overlap with the true ground state.

The area law has motivated several tensor-network \"ansatze to describe higher dimensional systems, in an attempt to replicate the success of DMRG. One example is the projected entangled-pair state (PEPS)~\cite{Nishino2001,Maeshima2001,Verstraete2004,Murg2007,Jordan2008,Gu2008,Jiang2008,GarciaSaez2011}, which can be thought of as a higher-dimensional generalization of MPS. The multi-scale entanglement renormalization ansatz (MERA)~\cite{Vidal2007b,Evenbly2009,Cincio2008} adds additional local entanglement to the TTN, and in its simplest form (c.f. \cite{Evenbly2012}) exactly replicates the area law in two- or higher-dimensional systems. The drawback of these approaches has been the large numerical cost, typically scaling as $\chi^{10}$ or greater. Fortunately, one expects that as computational power increases, the accuracy of these \"ansatze will increase superpolynomially with $\chi$ (for gapped phases).

In the mean-time, there is immediate demand for techniques with lower computational cost. Some approaches that have been tried recently include entangled plaquette states, and performing variational Monte Carlo over tensor network states~\cite{Schuch2008, Changlani2009,Mezzacapo2009,Marti2010,Sandvik2007,Wang2011,Ferris2011b,Ferris2011a}. Tensor network techniques that do not obey area laws have been used extensively in recent studies of 2D quantum systems, including DMRG in a cylindrical geometry~\cite{Yan2011,Stoudenmire2012,Depenbrock2012}, TTN~\cite{Tagliacozzo2009}, and direct approximate contractions of the 3D Suzuki-Trotter decomposition using tensor renormalization-group (TRG) and its variants~\cite{Levin2007,Gu2009,Xie2009,Zhao2010,Xie2012}. In these approaches, a description of a locally-correlated state would require a bond-dimension that grows with system size (e.g. exponentially with cylinder width in 2D DMRG).

There are two possible approaches to take in these cases: (a) use a small, finite geometry while keeping track of all correlations; or (b) study a large system using an ansatz with insufficient entanglement or bond-dimension $\chi$. The first approach is typically used because finite-size effects are well-understood, while finite-$\chi$ effects are less clear in the severely undersaturated regime. The purpose of this paper is to analyze, in generality, the scaling of accuracy in approach (b). One reason for this to be important is that approach (b) is implicitly used in the 3D classical tensor renormalization-group (or (2+1D) quantum TRG using a 3D representation of imaginary-time evolution) --- a promising technique that recently demonstrated accuracy competitive with large-scale Monte Carlo studies~\cite{Xie2012}.

We see that following a na\"ive approach, the global accuracy of (b) scales only logarithmically with $\chi$, and thus also logarithmic in the numerical cost.  On the other hand, provided proper care is taken, the accuracy of local observables scales polynomially with $\chi$. This exponential improvement puts the approach (at least formally) on similar grounds to finite-size scaling of small systems using DMRG. The `trick' here is to realize that the anisotropic structure of the renormalization means that not all sites are equal. Some sites are far away from the boundaries of the renormalization, and are able to share ample entanglement with their surrounding environment. In this paper, ideally located sites are called `center' sites.

The paper is structured as follows. In \Sec{sec:TTN}, the 2D tree-tensor network is analyzed in-detail, with arguments for the above scaling backed up by numerical results for the transverse-field Ising model on the square lattice. A similar analysis for the tensor renormalization-group in higher-dimensions is shown to hold in \Sec{sec:TRG}, with numerical evidence from the (classical) 3D Ising model (which is related to the (2+1)D quantum model). The paper concludes in \Sec{sec:conclusion} with an outlook on some possible future directions.

\section{Tree tensor network} \label{sec:TTN}

In this section we analyze the tree tensor network on an $L \times L$ square lattice. Here we use the 2-to-1 renormalization scheme depicted in \Fig{fig:TNs}~(c), alternating course graining along the $x$ and $y$ dimensions. The cost of contracting the tensor network corresponding to the expectation value of a nearest-neighbor Hamiltonian, $\langle \Psi | \hat{H} | \Psi \rangle$, scales as $\mathcal{O}(\chi^4 L)$, and similarly for calculating the derivative of the energy with respect to the tensor variables.

\begin{figure}
\begin{centering}
  \includegraphics[width=0.55\columnwidth]{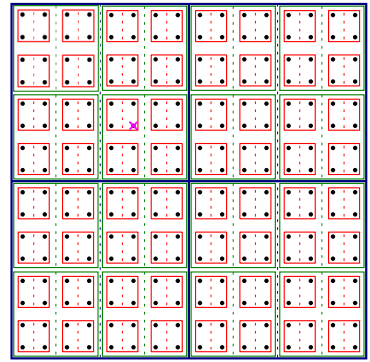}
  \caption{(Color online) Blocks of different layers in the 2D TTN. The marked `center' site at coordinates (6,6) is furthest from the block boundaries. This should be the site where the Hilbert space of its immediate environment is largest, and closest to the bulk, for all $\chi$. \label{fig:TTN}}
\end{centering}
\end{figure}

The wave function $| \Psi \rangle$ generated by the TTN contains blocks of size $l \times l$ (where $l = 2^n$, for some integer $n$) that contain bounded entanglement (Schmidt rank $\chi$) with the rest of the system. These blocks are highlighted in \Fig{fig:TTN}. In two-dimensions, this fails to saturate the area law, which demands that the entanglement entropy should scale linearly with the perimeter of the block, requiring a Schmidt rank scaling as $\exp(\alpha l)$, for some constant $\alpha$.

We now proceed to analyze the structure of the wave-functions having minimal energy. For large enough system size $L$, the bond dimension will be insufficient to describe the entanglement of the entire system, i.e. $\chi < \exp(\alpha L)$, and thus the minimum energy wave-function must be distinct from the true ground state. On the other hand, $\chi$ will be sufficient to describe the entanglement \emph{within} the smaller blocks. Errors will accumulate primarily because of the lack of entanglement \emph{between} larger blocks.

Somewhere between these two extremes there will be a critical block-size $l^{\ast}$, where smaller blocks have sufficient entanglement and are thus well-renormalized, whereas larger blocks do not possess large enough $\chi$. This is the point where
\begin{equation}
   \chi \sim \exp(\alpha l^\ast). \label{lstar}
\end{equation}

Because the blocks grow exponentially in size as a function of layer $n$, and the required bond-dimension therefore grows doubly-exponentially in $n$, we expect the transition between sufficient entanglement and woefully inadequate $\chi$ to be quite sharp.

To simplify the analysis, we compare this situation to a cluster-mean field theory. In this theory, the full Hilbert space of clusters of size $l^{\ast} \times l^{\ast}$ is included, while no entanglement exists between neighboring clusters. In practice, a single cluster is exactly-diagonalized in a self-consistent fashion with their boundary conditions, and the fixed-point corresponds to the lowest energy state in the cluster mean-field ansatz.

The error of global quantities such as the total energy can be reasonably large when using this approach. Let's assume that the system has correlation length $\xi$, and that the expectation value of a local quantity decays exponentially towards the (correct) bulk value away from the cluster boundaries. In the limit that $l^{\ast} > \xi$, some fraction of the system will be `close' to the boundaries (closer than $\xi$) and display incorrect results, while the remaining `bulk' fraction will display roughly the correct results. In $d$-dimensions, the fraction of `error' sites scales as  $2d \xi / l^{\ast}$, and thus the error of a global quantity in the 2D TTN scales as
\begin{equation}
   \text{global error}  \propto \frac{ \xi}{l^{\ast}} \sim \frac{\xi \alpha}{ \ln \chi}
\end{equation}
where \Eqn{lstar} was used in the second relation. The error of a global quantity, such as the total energy or magnetization, thus only decreases logarithmically with $\chi$ and thus computation effort.

Although this scaling is quite poor, the ansatz takes some advantage from the localized entanglement in the system and is already a large improvement over exact diagonalization\cite{Tagliacozzo2009}. On top of this, we can estimate \emph{local} quantities with greatly reduced error by understanding the structure of the ansatz. Points inside the well-renormalized, `bulk' region are surrounded by an immediate environment that is a good approximation of the true ground state (provided, again, that $\l^{\ast} > \xi$). As mentioned earlier, in a gapped phase the effect of the boundary should decay exponentially, so in the centre of the region the error should scale as $\exp(-l^{\ast}/2\xi)$. This time, including \Eqn{lstar} gives
\begin{equation}
   \text{local error}  \sim \exp \left( \frac{-l^{\ast}}{2\xi} \right) \sim \exp \left( \frac{-\ln \chi}{ 2 \xi \alpha} \right) \sim \chi^{-\beta} ,
\end{equation}
for some $\beta$ that depends on the specifics of the system (notably, becoming smaller for more entangled systems or those with larger correlation length). Thus the scaling of error with computation effort is \emph{polynomial} --- an exponential improvement on the error of global quantities. If $\beta$ is very large, the method could become competitive with well-established techniques such as quantum Monte Carlo (where statistical error scales as the square root of computational effort, assuming no sign-problem). The performance will degrade significantly in systems with large amounts of entanglement, or close to critical points.

To investigate the above numerically, we identify points in the system that are somehow `central' to the renormalization, independent of system parameters or $\chi$. In the TTN, these are the points that are renormalized best with their environment, favoring no particular direction. In the 2D TTN, we define these to be the points that have been successively combined with the block above, to the left, below, to the right, \emph{ad infinitum} (see \Fig{fig:TTN}). For any choice of $\chi$, such a point has the largest immediate environment whose sites are described by a sufficiently large Hilbert space, roughly equal in all directions.

We have implemented the 2D TTN to study the spin-1/2 transverse-field Ising model on the square lattice, described by Hamiltonian
\begin{equation}
  \hat{H} = -\sum_{<i,j>} \hat{\sigma}^z_i \hat{\sigma}^z_j + h \hat{\sigma}^x_i, \label{TFI}
\end{equation}
where $<\!\!i,j\!\!>$ denotes neighboring sites and $h$ is the strength of a transverse magnetic field. Because the cost of the algorithm cost scales as $\mathcal{O}(\chi^4 L)$, a moderate size $L=32$ is used in order to investigate a system that is much larger than can be described exactly, while small enough to allow a range of $\chi$ to be used. To minimize the energy, I have employed the time-dependent variational principle~\cite{Haegeman2011} and used a unitary gauge, finding a significant speed-up compared to the traditional, SVD approach~\cite{Tagliacozzo2009}.

\begin{figure}
\begin{centering}
  \includegraphics[width=\columnwidth]{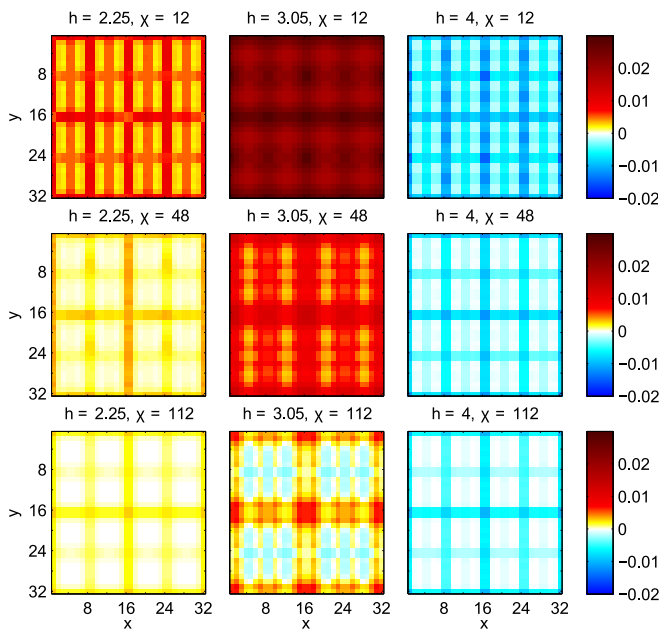} 
  \caption{(Color online) Spatial variations in the magnetization in field direction at various values of $h$, as predicted by the lowest-energy TTN with the displayed bond-dimensions $\chi$. The color values represent differences to the prediction at $\chi=112$. In the gapped regions away from the critical point (near $h=3.05$), we observe a `windowing' pattern, with window size growing with $\chi$, and rapid convergence at the central points of the renormalization (at $x,y=21$). When the correlation length is longer, the effect is blurred somewhat and convergence was not reached with a value of $\chi = 112$. \label{fig:xmag}}
\end{centering}
\end{figure}

In \Fig{fig:xmag} the error of the magnetization in the $x$ direction (as compared to the best, center site estimate) is displayed. We can clearly observe a pattern of successive `windows', where errors are displayed primarily on the \emph{edges} of blocks. For larger $\chi$, the size of the windows grow while the overall error decreases --- while the central sites always remain inside the smallest window. At the critical point, close to $h=3.05$, there is a large correlation length and the windows are somewhat blurred, and the errors are greater in magnitude for a given $\chi$.

\begin{figure}
\begin{centering}
  \includegraphics[width=\columnwidth]{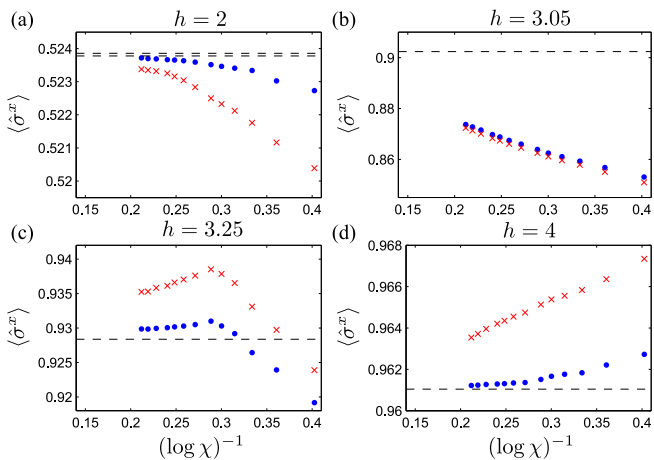}
  \caption{(Color online) Predictions for the magnetization in the field direction with various magnetic field strengths $h$, as predicted by the lowest-energy TTN with given bond-dimension $\chi$. The red crosses correspond to system-wide averages, while the blue points correspond to the central sites of the renormalization. The latter converge much faster than the former, away from the critical value of $h\approx 3.05$. The dashed lines correspond to QMC results for the system at temperature $T=0.005$. In (a) the two lines represent the statistical uncertainty, while in (b--d) the error is comparable to the line width. \label{fig:xplots}}
\end{centering}
\end{figure}

We compare the predicted value of $\langle \hat{\sigma}^x \rangle$ from the global, average value and the local value at the center site in \Fig{fig:xplots}. Away from criticality, we observe that the center site value converges to the quantum Monte Carlo prediction (using ALPS~\cite{ALPS20,ALPS13}) significantly faster than the global average. At criticality, both values appear to be converging with a slow, $1/\ln \chi$ scaling (extrapolation may be viable when $L$ approaches the thermodynamic limit). At $h=3.25$ the behavior is not monotonic because the wave-function switches from the ferromagnetic to disordered phase at intermediate $\chi$.

These results confirm the above analysis and show that it is viable to extract meaningful information in systems where the parameter $\chi$ is severely undersaturated.

\section{Tensor renormalisation group} \label{sec:TRG}

\begin{figure}
\begin{centering}
  \includegraphics[width=0.9\columnwidth]{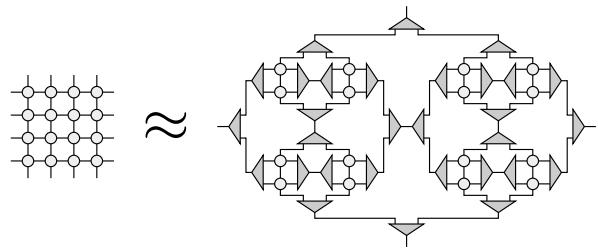}
  \caption{The contraction of the 2D Markov network (i.e. partition function) on the left is approximated by the TRG scheme on the right, from~\cite{Xie2012}. In the `second renormalisation group', the projectors (triangular tensors) are optimized to maximize the partition function. The above diagram can also be considered an ansatz for the probability distribution itself, by opening additional legs on the Markov network tensors (circles) corresponding to the state at that site.  \label{fig:TRG}}
\end{centering}
\end{figure}

A related tensor-based, real-space renormalization technique is the tensor renormalization group (TRG). In this family of methods, the tensor network corresponding to the Markov network of a classical thermal state (whose contraction gives the partition function) is successively contracted into larger and larger blocks. At each layer, the dimension of the tensors must be truncated to prevent the difficulty from growing exponentially. The process is depicted in \Fig{fig:TRG}.

The method is well-suited to contracting 2D tensor networks, such as thermal state of a 2D classical system, or the network that results from the Trotter decomposition of a 1D quantum density matrix. For non-critical systems, the fixed-point of this RG flow is well-understood~\cite{Gu2009} in terms of `corner' correlations. The TRG, and related `second renormalization group' (SRG) methods are able to encapsulate the correlations of the 2D classical system quasi-exactly with fixed $\chi$.

This technique has been successfully applied in a 3D classical (or 2D quantum) setting, achieving impressive accuracies (see e.g. Ref.~\cite{Xie2012}).  However, in these higher-dimensional settings, the edges of the 3D blocks grow in length as the renormalization proceeds, and strictly speaking $\chi$ would need to grow exponentially to encapsulate all the correlations in a (non-critical) system. In general, for this kind of blocking scheme in $d$ dimensions, the correlations that need to be to be accounted for at each level of the renormalisation grows as $L^{d-2}$. Thus, the TRG fails to account for what one might call a `corner' law for correlations~\cite{GuifrePrivate,Riera2013}, related to the quantum area law for entanglement entropy that manifests naturally in a $(d+1)$-dimensional quantum theory, where one dimension is time.

\begin{figure}
\begin{centering}
  \includegraphics[width=0.5\columnwidth]{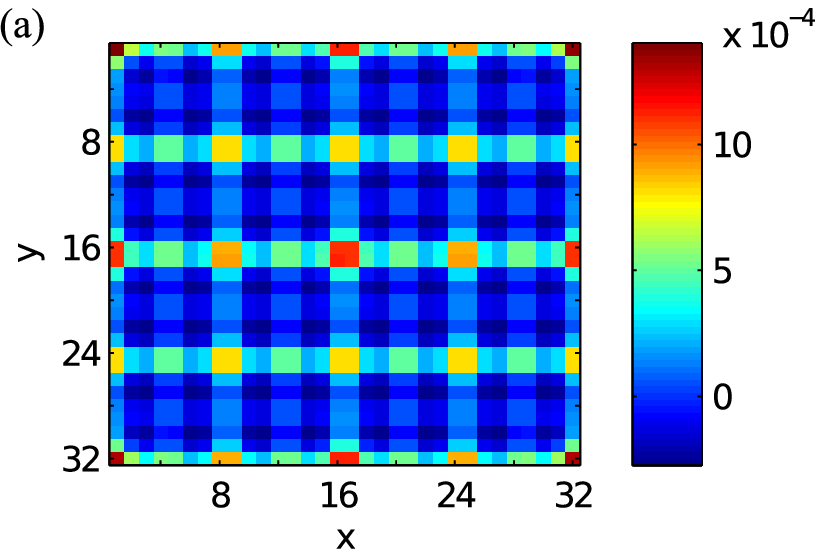}\includegraphics[width=0.5\columnwidth]{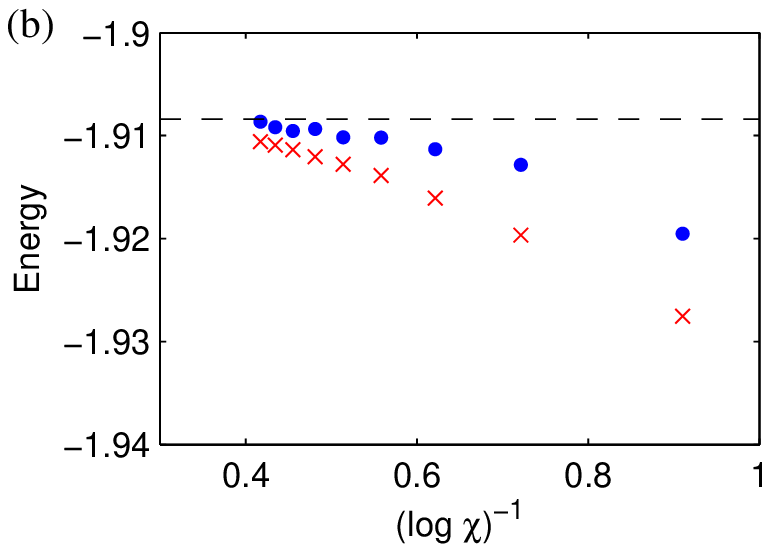}
  \caption{(Color online) Predictions for the 3D Ising model at $T=4$ (less than the critical temperature of $T\approx 4.1$) using the HOSRG approach. (a) The bond-energy of the $z$-bonds through a cross-section of the $x$--$y$ plane, where the value at the center site labeled $x,y=21$ has been subtracted. The 3D renormalization scheme displays errors predominantly on the corners of the renormalized regions. (b) The energy (per-site) for different values of $\chi$, compared to classical Monte Carlo results. The red crosses correspond to system-wide averages, while the blue points correspond to the central sites of the renormalization. Similar to the quantum TTN, the latter converge much faster than the former. The dashed line corresponds to Monte Carlo simulations of a 48$\times$48$\times$48 lattice. \label{fig:srgplots}}
\end{centering}
\end{figure}

The structure of the TRG is strikingly similar to the TTN and one might suppose that a similar analysis in the limit of large 3D systems with insufficient $\chi$ might hold. Numerical evidence in the 3D Ising model appears to support this claim. I implemented the 3D higher-order SRG (HOSRG) approach from Ref.~\cite{Xie2012}, using an SVD-update~\cite{Tagliacozzo2009,Evenbly2007} to optimize the tensors to maximize the (global) partition function. The computational effort scales as $\mathcal{O}(\chi^{11} \ln L)$, and we are able to study much larger systems, with $L = 2^{12}$. In \Fig{fig:srgplots}~(a) we see a clear indication of the `corner' errors in the local energy, in a similar fashion to the `window' pattern in \Fig{fig:xmag}. In \Fig{fig:srgplots}~(b) the average energy and center site energy are compared for different values of $\chi$, and we see even close to the critical point (at $T\approx4.1$) the local quantity converges to the Monte Carlo prediction much more rapidly.

\section{Discussion} \label{sec:conclusion}

We have analyzed real-space renormalization procedures that are unable to account for local correlations in higher-dimensional quantum and classical systems. The anisotropy of tree-structured \"ansatze, such as TTN and TRG, can be taken advantage-of to identify sites that are central to the renormalization, providing an (exponentially more) accurate description of local quantities. The scaling of accuracy to numerical cost is expected to be polynomial (for gapped systems), formally putting the technique on a similar footing to 2D DMRG of small systems and quantum Monte Carlo (though in practice the algorithms used here might not be as efficient).

There has been some reluctance to use a severely under-correlated ansatz for a quantum wave-function, where recent focus has been on DMRG in small geometries such as narrow cylinders. On the other hand, the TRG approach has shown very promising results in 3D classical (and 2D quantum) systems, while there has been less discussion on the inherent inability to account for all correlations in a large system. In-fact, it is possible that the central-site technique has been implemented in these studies in the past.

From here, two possible directions to increase the effectiveness of real-space RG techniques in higher dimensions become apparent. The first would be find more efficient algorithms in the under-correlated regime. For example, the HOSRG was a step in this direction~\cite{Xie2012}, compared to earlier 3D TRG algorithms. In the present work, it could be beneficial to replace the TTN with a MPS having a tree-like structure, investigated already in Ref.~\cite{Xiang2001}. The cost would reduce to $\mathcal{O}(\chi^3 L^2)$, but one would still be limited to moderate system sizes.

The second approach would be to use an ansatz that takes into account the correlation structure of the system. PEPS and MERA already exist to describe higher-dimensional quantum systems. In the realm of classical Markov networks, progress was made in Ref.~\cite{Gu2009} to use a TRG-like approach to take account of all local correlations (for 2D classical systems). However, a more direct, MERA-like approach to 2D and 3D classical systems would represent a major advancement in this field.

\subsection*{Acknowledgements}

I would like to thank Guifre Vidal and David Poulin for discussions. This work was supported by NSERC and FQRNT through the network INTRIQ, as well as the visitor programme at the Perimeter Institute for Theoretical Physics. 

\bibliography{../bib/andy}

%
%

\end{document}